\documentclass[%
 reprint,
superscriptaddress, 
 amsmath,amssymb,
 aps, prl, floatfix
]{revtex4-2}
\usepackage[utf8]{inputenc}
\usepackage[T1]{fontenc}
\usepackage{csquotes}
\usepackage{silence}
\usepackage{pgffor}
\newcommand{\vect}[1]{\boldsymbol{#1}}

\usepackage{xcolor} 
\usepackage[colorlinks=true,citecolor=blue,linkcolor=blue,urlcolor=blue]{hyperref}

\usepackage{pdfpages}
\usepackage{pgffor}

\makeatletter
\AtBeginDocument{\let\LS@rot\@undefined}
\makeatother
\begin{document}

\preprint{}


\title{Collective Dynamics in Active Polar Polymer Assemblies}

\author{Hossein Vahid}
\affiliation{Leibniz-Institut f{\"u}r Polymerforschung Dresden, Bereich Theorie der Polymere, 01069 Dresden, Germany}

\author{Jens-Uwe Sommer}
\email{jens-uwe.sommer@tu-dresden.de}
\affiliation{Leibniz-Institut f{\"u}r Polymerforschung Dresden, Bereich Theorie der Polymere, 01069 Dresden, Germany}
\affiliation{Technische Universit{\"a}t Dresden, Institut f{\"u}r Theoretische Physik, 01069 Dresden, Germany}

\author{Abhinav Sharma}
\email{abhinav.sharma@uni-a.de}
\affiliation{Faculty of Mathematics, Natural Sciences, and Materials Engineering: Institute of Physics, University of Augsburg, Universit{\"a}tsstraße 1, 86159 Augsburg, Germany}
\affiliation{Leibniz-Institut f{\"u}r Polymerforschung Dresden, Bereich Theorie der Polymere, 01069 Dresden, Germany}

\begin{abstract}
{Tangentially driven active polymers (TDAPs), model systems for motor-driven filaments, have been extensively studied in uniform activity fields.
Here, we show that an activity gradient breaks fore–aft symmetry, generating net body forces that steer dimers, asters, and larger assemblies toward high-activity regions.
Including temporal stochasticity softens the chains, allowing them to bend and wind around other filaments.
Once several contacts are established, steric interlocking arrests relative motion and stabilizes the assembly into a hierarchically entangled cluster.
These clusters persist for times far exceeding single-chain relaxation and do not appear under deterministic, temporally constant activity. Remarkably, such activity-induced gelation occurs even at polymer concentrations substantially lower than those typically required for passive chains.
Our results reveal a new mechanism for activity-induced aggregation, providing new strategies for designing autonomous and reconfigurable microfluidic systems.}
\end{abstract}

\date{\today}

\maketitle
\textit{Introduction.}---{Polymer chains propelled tangentially along their backbone provide a framework for studying emergent behavior of active worm-like chains and motor-driven filaments~\cite{mokhtari2019,sinaasappel2025,schaller2010,de2017,anand2018,bianco2018,philipps2022,li2023,fazelzadeh2023,karan2024,khalilian2024,vatin2024, ubertini2024}.
Previous studies have largely focused on weakly interacting polymers in homogeneous and time-independent activity fields, where individual chains evolve from straight to spiral or globular conformations as activity increases~\cite{isele2015, bianco2018, zhao2024, karan2024, khalilian2024, vansteijn2024}, and collectively, transition from jamming at low activity to nematic order at intermediate levels, and active turbulence at high activity~\cite{duman2018}.
However, biological systems often feature spatially and temporally varying energy input, leading to pronounced structural complexity, including aggregation, entanglement, and gelation~\cite{prost2015, schuppler2016, demir2020, deblais2023, zhang2021}.
}

{
We demonstrate that spatial activity gradients and temporal propulsion noise induce the formation of non-equilibrium entangled structures between assemblies of tangentially driven active polymers (TDAPs) at concentrations well below those required for passive polymers.
Single TDAPs accumulate in low-activity regions without entangling.
In contrast, activity gradients bias outward-driven dimers, asters, and larger assemblies toward high-activity regions, and the temporal stochasticity in propulsion increases conformational flexibility, allowing polymers to bend, loop, and interlock. The resulting aggregates are stabilized purely by entanglements, without any adhesive bonds or attractive interactions.
Compared with spatially uniform activity, an activity gradient accelerates entanglement nucleation and increases aggregate persistence.
}

\begin{figure}[b!]
    \centering
    \includegraphics[width=0.8\linewidth]{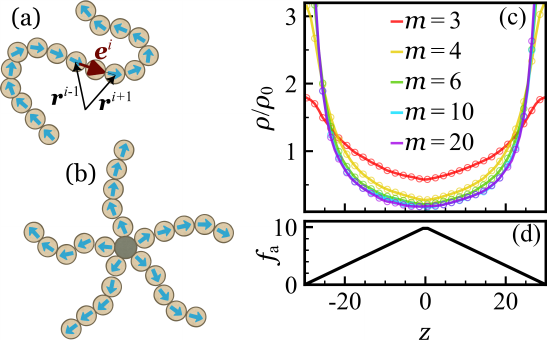}
    \caption{Cartoon representation of (a) a TDAP and (b) an aster. Each active monomer is self-propelled in the direction of the local tangent to the backbone as indicated by the blue arrows. (c) Steady-state density of monomers along the $z$ axis for varying degrees of polymerization $m$. The bulk monomer density is $\rho_0=0.002$. Polymers are simple TDAPs as in (a), and the activity field is given by $f_{\rm a} = 10 ( 1 - |z|/30)$ {(see panel (d))}. TDAPs accumulate in the low-activity regions. 
}
    \label{fig1}
\end{figure}

\textit{The model.}---We perform Brownian dynamics simulations to investigate semiflexible TDAPs and multi-arm TDAPs, see Fig.~\ref{fig1}(a,b).
Each TDAP monomer experiences a self-propulsion force $\vect{F}^i_{\textrm{a}} = f_{\rm a} \vect{e}^i$, where $f_{\rm a}$ is the magnitude of the active force, and $\vect{e}^i$ denotes the unit vector indicating the propulsion direction.
The propulsion direction of each monomer (except for the polymer ends) at position $\vect{r}^i$ is aligned with the local tangent vector of the polymer backbone, updated at each time step by $\vect{e}^i(t)=\vect{t}^i/\vert \vect{t}^i\vert$, where $\vect{t}^i=\vect{r}^{i+1} - \vect{r}^{i-1}$, and $\vect{r}^{i-1}$ and $\vect{r}^{i+1}$ represent the positions of the adjacent monomers.
For the first and last monomers of the chains, $\vect{e}^i$  aligns with the bond connecting them to their nearest neighbor monomers.

The equation of motion of each monomer is described by $\gamma_{\textrm{t}}\dot{\bm r}^i = -\sum_j\nabla_{\bm{r}^i}U^{ij} + \bm{F}_{\textrm{a}}^i + \bm{\xi}^i(t),$
where $\gamma_{\textrm{t}}$ is the translational friction coefficient of particle $i$, and $U$ the potential energy.
The stochastic noise $\vect{\xi}^i(t)$ is Gaussian, with zero mean $\langle \vect{\xi}^i(t) \rangle = 0$ and autocorrelation function $\langle {\xi}^i_\alpha(t). {\xi}^j_\beta(t') \rangle = 2 \gamma_{\rm{t}}^{-1} k_{\rm{B}} T \delta^{ij} \delta_{\alpha\beta} \delta(t-t')$, where $k_{\rm B}$ is the Boltzmann constant, $T$ the temperature, and $\alpha, \beta \in \{x, y, z\}$.
Interparticle interactions are modeled using the Weeks-Chandler-Andersen (WCA) potential~\cite{anderson}, $U_{\rm WCA}^{ij}(r) =
        4\epsilon [ ( \frac{\sigma^{ij}}{r^{ij}})^{12}- (\frac{\sigma^{ij}}{r^{ij}})^{6}+\frac{1}{4} ]\Theta(r_{\rm c}^{ij}-r^{ij}),$
where $r^{ij}$ is the distance between particles $i$ and $j$, and $\sigma^{ij}=0.5(\sigma^i+\sigma^j)$ is their effective interaction diameter, with $\sigma^i$ being the diameter of particle $i$. Here, $\epsilon$ is the depth of the potential well, $\Theta$ is the Heaviside step function, and the cutoff radius is set to $r_{\rm c}^{ij}=2^{1/6}\sigma^{ij}$.
In addition to WCA interactions, the bonded monomers are connected using the finite extensible nonlinear elastic (FENE) potential~\cite{bird1987}, defined as $
U_{\rm F}(r) = 
-\frac{1}{2} k_{\rm F} R_0^2 \ln [ 1 - ( \frac{r^{ij}}{R_0} )^2 ]\Theta(R_0-r^{ij}),
$
where $k_{\rm F}$ represents the elastic coefficient and $R_0$ is the maximum bond length. 
The chain conformation is controlled by the bending potential $U_{\rm b}= k_{\rm b} \left[ 1 - \cos(\theta -\theta_0)\right]$,
where $k_{\rm b}$ is the bending modulus, $\theta$ the angle between consecutive bonds, and $\theta_0$ the rest angle.

We set $\sigma = 1$, $\epsilon = k_{\rm B}T = 1$, and $\tau = \sigma^2 \gamma_{\rm t}/(3k_{\rm B}T)$, with $\gamma_{\rm t}=3$, as the units of length, energy, and time, respectively.
All simulations are conducted using the \textsc{lammps} package~\cite{plimpton1995, Thomson2022} within a cubic simulation box of dimensions $60\times60\times60\sigma^3$, ranging from $-30$ to $30$ along each axis. Periodic boundary conditions are imposed in all directions. The activity field varies linearly along the $z$-axis according to
$f_{\rm a}(z) = f^*_{\rm a} ( 1 - |z|/30)$,
where \( f_{\rm a}=f^*_{\rm a} \) at the box center $z=0$, and $f_{\rm a}=0$ at $\vert z \vert=30$.
{In a subset of simulations, we impose temporal stochasticity on the activity field.
To this end, the box is partitioned along $z$ into slabs of width $0.5$.
At each time step, an independent uniform random variate $\eta(z,t)\in[0,1]$ is drawn for each slab, and the local activity is updated as $f_{\rm a}(z,t)=f_{\rm a}(z)\,\Theta\!\big[P-\eta(z,t)\big]$. Here, \(P\in[0,1]\) is a control parameter setting the activation probability (mean duty cycle); thus, the local activity of the slab equals $f_{\rm a}(z)$ with probability $P$ and $0$ otherwise.}
The simulation parameters are set to $\sigma^i=1$, $k_{\rm F}=30$, $R_0=2\sigma^{i}$, $k_{\rm b}=30$ and $\theta_0=120^\circ$.
Additional computational details are described in the Supplementary Material (SM).

\textit{Results.}---We first study individual TDAP chains in an activity field given by $f_{\rm a}=10(1 - \vert z \vert/30)$. Fig.~\ref{fig1}(c) shows that these polymers tend to accumulate in low-activity regions, regardless of the chain length.
TDAP propulsion aligns with the polymer backbone, and thus, following the motions of their head monomers, TDAP chains preferentially orient and accumulate toward lower-activity regions.
{Including temporal stochasticity does not alter the single-arm TDAP response. They still accumulate in low-activity regions, with only a slight increase as $P$ increases (cf. Fig.~S3 of Supplementary Material (SM)).}
{In contrast, active Brownian polymers (ABPOs) exhibit length-dependent accumulation, and migrate to high-activity regions as the degree of polymerization, $m$, increases~\cite{vuijk2021,ravichandir2024,muzzeddu2024,valecha2025}; cf. Fig.S1 of SM. Monomeric ABPs and dimers favor low-activity regions, and at sharp motility gradients, Janus microswimmers polarize from high to low activity so that their density in regions is set by the local activities~\cite{soker2021,auschra2021,vuijk2021}.
Temporal switching does not qualitatively change the ABPO response. At very low $P$, the mean propulsion is too weak to sense the gradient, reducing drift and accumulation in the high-activity region (Fig.~S3, SM).}

\begin{figure}[b!]
    \centering
    \includegraphics[width=0.8\linewidth]{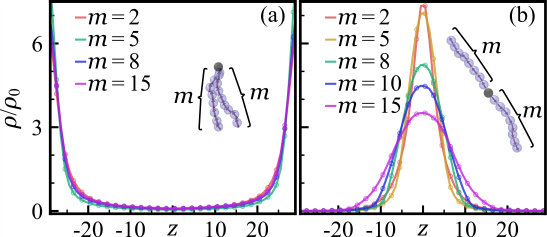}
    \caption{Steady-state density of monomers along the $z$ axis. Polymers have two symmetric TDAP arms, each containing $m$ monomers, connected to a passive core, with $\rho_0=0.002$. $f_{\rm a}(z)$ follows that of Fig.~\ref{fig1}(d). (a) TDAPs are directed toward the core and accumulate in the low-activity regions. (b) TDAPs are directed away from the core and migrate to high-activity regions.}
    \label{fig2}
\end{figure}

\begin{figure}[t!]
    \centering
    \includegraphics[width=1\linewidth]{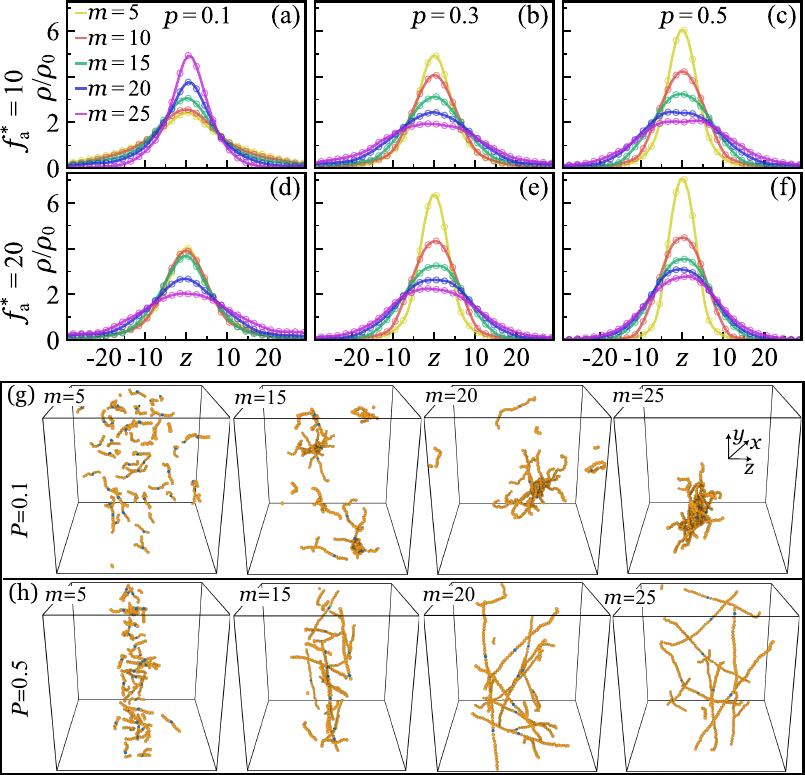}
    \caption{{Steady-state monomer density along the $z$-axis for varying $P$ and $m$ for two-arm outward-directed TDAP assemblies. The activity field is given by $f_{\rm a}=f_{\rm a}^*\left(1-|z|/30\right)\Theta\left[P - \eta(z,t)\right]$;(a-c) $f_{\rm a}^*=10$, (d-f) $f_{\rm a}^*=20$. Bulk density of monomers is $\rho_0=0.002$.
    Each polymer has two symmetric arms ($m$ monomers each) attached to a passive core. (g) and (h) show representative snapshots of the systems in (a) and (c), respectively. TDAPs are colored yellow, and the cores are blue. Movies~1–4 of SM show the simulation animations at \(f_{\rm a}^*=5,10, {\textrm{ and }}20\).}}
    \label{fignew}
\end{figure}

In many biological systems, active filaments are observed in assembled structures, such as bundles and asters~\cite{nguyen2014, ishihara2014, feng2017, woodruff2017, gupta2017, norris2018, soares2011, huber2012, bovellan2014, fritzsche2017, colin2019-actin}.
Figure~\ref{fig2} shows that the directed motion and the overall conformation of two connected TDAPs are significantly influenced by their relative propulsion directions. 
We connect two TDAP arms, each consisting of $m$ monomers, to a central passive core. We consider two distinct configurations: inward-directed (tangential propulsion from the arm tips toward the core, Fig.\ref{fig2}(a)) and outward-directed (from the core toward the arm tips, Fig.\ref{fig2}(b)).
In the inward-directed case, the two arms tend to approach each other, which again leads to accumulation in lower-activity regions. Inward propulsion creates effective inward-directed stresses leading to arm folding at the core position that stabilizes their collective migration toward low activity. Variations in $m$ do not significantly influence this behavior.

For outward-directed propulsion, the two arms exhibit a tug-of-war dynamics~\cite{muzzeddu2022}.
The tug-of-war of motor proteins was also found to be highly cooperative and perform directed cargo transport~\cite{muller2008}.
Here, tangential propulsion extends the chain, and the arm sampling the higher-activity side experiences the larger thrust; the resulting net force points up the gradient, leading to drift and accumulation in the high-activity region.
{This behavior corresponds to rigid ABP dimers exhibiting polarity-dependent accumulation. In this case, it was shown analytically that the outward-oriented dimers migrate to high-activity regions, whereas the inward-oriented dimers accumulate in low-activity regions~\cite{vuijk2022}.}
Figure~\ref{fig2} shows that as arm length increases, monomers tend to accumulate in high-activity regions more, but their spatial distribution becomes broader, and the peak density at the highest activity decreases. Since the polymer remains nearly stretched, not all monomers can occupy the same $z$ position. While the core remains in the high-activity region (see Fig.~S6 of SM), the extended arms reach lower-activity areas.
Furthermore, our results show that as the polymer becomes more stretched (increasing $\theta_0$) or stiffer (increasing $k_{\rm b}$), the accumulation in the high-activity regions enhances (cf. Fig.~S8 of SM).
In biological systems, crosslinked actin and microtubule (MT) structures often exhibit asymmetric arm lengths~\cite{schuppler2016}. Increasing asymmetry eventually leads to accumulation in low-activity regions. 
Interestingly, at intermediate asymmetry, polymers accumulate in the intermediate activity regions due to the competition between arms (cf. Fig.~S9 of SM).

{Figure~\ref{fignew} demonstrates how stochastic activity (activation probability $P$) and the maximum propulsion strength $f_{\rm a}^*$ control the collective organization and spatial distribution of two-arm, outward-directed TDAP assemblies.
For short chains (e.g., $m=5$), increasing either $f_{\rm a}^*$ or $P$ enhances the tug-of-war between the arms, leading to enhanced accumulation in high-activity regions. 
Unexpectedly, long chains (e.g., $m=25$) at low $P$ and low $f_{\mathrm a}^{*}$ become flexible and entangle into stable, aster-like aggregates (panel g).
The longer the chain, the more stable the entangled structures (cf. Movie~1-4 of SM).
Increasing $P$ or $f_{\mathrm a}^{*}$ stiffens the arms, suppresses entanglement, and permits extended conformations, so the peak density decreases.
Bond-vector autocorrelations confirm this trend, decaying more rapidly for small $P$ (cf. Fig.~S7).
No persistent entanglement is observed in the deterministic limits $P=1$ (always on) or $P=0$ (always off).
Once steady state is established, inverting the gradient (i.e., $f_{\mathrm a}(z)=10|z|/30$) translates the pre-formed aggregate toward the new maximum; transiently detached filaments re-associate in the high-activity region (Movie~4, SM).
 }


\begin{figure}[b!]
    \centering
    \includegraphics[width=0.8\linewidth]{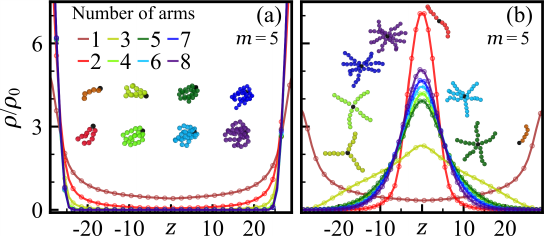}
    \caption{
    Steady-state density of monomers along the $z$ axis for polymers with varying numbers of arms. Each arm is an (a) inward- or (b) outward-directed TDAP with 5 monomers.
    The activity field is given by $f_{\rm a} = 10 ( 1 - |z|/30)$, and the bulk monomer density is set to $\rho_0=0.002$.
    }
    \label{fig3}
\end{figure}

Figure~\ref{fig3} presents multi-arm stars, each containing five monomers ($m=5$) extending from a central passive core. 
{Such multi-arm geometries are common in cytoskeletal networks and cellular assemblies~\cite{weber2015,brangwynne2008}.
Recent experiments showed control over inward- and outward-directed forces on MTs using light-switchable motors~\cite{lemma2023} or by balancing motor activity and polymerization~\cite{schuppler2016}. Mixed active-passive polymers have also been realized experimentally in kinesin-driven MT-actin composites~\cite{berezney2022}.}
We examine inward-directed propulsion (toward the central core, Fig.~\ref{fig3}(a)) and outward-directed propulsion (away from the core, Fig.~\ref{fig3}(b)).
For inward-directed propulsion, arms tend to collapse close to each other, forming compact, bundle-like structures. These arms cooperatively move toward the low-activity region, and increasing the number of arms enhances accumulation in that region.

For outward‑directed propulsion, the polymer unfurls into an aster that migrates toward the high-activity region, yet the drift strength is nonmonotonic in arm number.
Maximum accumulation in the high-activity region is achieved for the two-arm structure, which decreases dramatically when a third arm is added.
Here, the three-arm polymer faces competition among arms oriented toward different activity levels, reducing the net directional migration. Increasing the number of arms to four slightly improves accumulation relative to the three-arm structure. Further increases in arm number have minimal effects.
These trends are robust to variations in $f_{\rm a}^*$ and bulk densities of monomers ($\rho_0$) (see Fig.~S13 of SM). At very high densities, accumulation extends to lower-activity regions because the high-activity regions become saturated.
{
For longer arms ($m>15$), the density profiles become nearly arm-number independent (Fig.~S12, SM).
Multi-arm ABPOs also accumulate in high-activity regions, and both increasing the number of arms and $m$ enhance their accumulation (Fig.~S12, SM).
We note that two outward-directed arms are sufficient to bias migration toward the high-activity region, irrespective of the polarity of the remaining arms (cf. Fig.~S11 of SM).}

\begin{figure}[t!]
    \centering
    \includegraphics[width=1\linewidth]{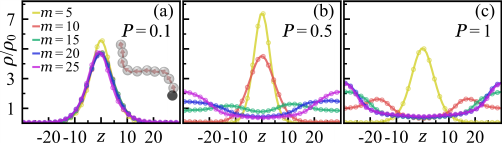}
    \caption{{Steady-state monomer density along the $z$-axis for varying $P$ and $m$. The activity field is given by $f_{\rm a}=5\left(1-|z|/30\right)\Theta\left[P - \eta(z,t)\right]$. Each polymer consists of a TDAP pulling an attractive cross-linker particle. Panel (a) includes an overlaid schematic of a TDAP with \(m=10\) (gray spheres), pulling a cross-linker (black sphere).
     Movie~5 of SM illustrates animations of simulations at varying $m$ and $P$.}}
    \label{fig5}
\end{figure}

{Cytoskeletal assemblies can emerge spontaneously, for example, through motor-driven reorganization, filament buckling, and cross-linking interactions~\cite{hentrich2010,gordon2012,tanenbaum2013,norris2018,tan2018,henkin2022,najma2024,utzschneider2024,lamtyugina2024,lemma2022,prost2015,tee2015,matsuda2024}. To simulate dynamic assemblies, we consider TDAPs that each pull a passive cross-linker attached at the tail. 
The cross-linkers interact attractively with one another via a Lennard-Jones potential with well depth $\epsilon=15$ (see Eq.~(S3) of SM), and all other interactions are described by the WCA potential, thus forming temporal asters with outward-directed activity of the arms.
The activity field is $f_{\rm a}(z,t)=5\!\left(1-|z|/30\right)\Theta\!\big[P-\eta(z,t)\big]$.
Figure~\ref{fig5} presents the steady-state monomer density profiles for varying $m$ and $P$. At low $P$ (e.g., $P=0.1$), attraction dominates over active forces and TDAPs form asters that accumulate in high-activity regions (cf. Movie~5 of SM). At $m=5$, many small asters appear, and increasing $m$ yields fewer, larger aggregates.
At the largest chain length considered ($m=25$), a single large aggregate forms, with individual TDAP chains dynamically exchanging between the aggregate and the surrounding region. The aggregates accumulate in the high-activity regions (cf. Movie~5 of SM).}

{
As $P$ increases, the competition between the chain activity and cross-linker attraction determines the structural stability and accumulation pattern of TDAPs.
Short chains ($m=5$) remain largely aggregated in asters across $P$, with slightly reduced stability at $P=1$. Thus, accumulation in the high-activity region slightly decreases at $P=1$.
Longer chains ($m\!\ge\!10$) are more sensitive to activity.
At intermediate $P$ (e.g., $P=0.5$), polymers with ($m=10$) still form stable asters, allowing pronounced accumulation in high-activity regions.
Further increasing $P$ leads to the formation of transient structures: asters initially assemble and move toward higher activity, but active forces eventually dominate, causing them to dissociate. This results in density peaks at intermediate positions along the gradient. For even longer polymers, attractive forces fail to stabilize aggregates at high $P$, resulting in the dispersal of individual chains to low-activity regions.}

To confirm the generality and robustness of our findings, we systematically explored how varying polymer conformation, activity magnitude, polymer concentration, the degree of partial activation, {and confinement} affect the observed behaviors (see SM). Across these variations, our results remain consistent, demonstrating the generality of the observed behaviors. 


\textit{Conclusions.}---{In uniform activity, directed transport of active polymers is strongly coupled to polymer configuration and structure~\cite{zhao2024,isele2015,karan2024,khalilian2024,vansteijn2024, ravichandir2024}.
Compact structures exhibit reduced mobility, whereas elongated configurations exhibit enhanced diffusion and directed motion~\cite{bianco2018,khalilian2024, jain2022}.
The coupling can be tuned by where activity is placed along the backbone and even by partial activation, which can induce knots~\cite{vatin2025} and improve transport efficiency~\cite{vatin2024}.
We have shown that stochastic activity enhances the accumulation of long polymer assemblies ($m>20$) in high-activity regions.
}

{In this Letter, we have demonstrated that spatial and temporal variations in activity can fundamentally transform the collective behavior of TDAPs.
Our findings show that (i) structured assemblies with at least two outward-driven arms robustly migrate to high-activity regions, regardless of whether the remaining arms are passive, inward-, or outward-driven.
(ii) Introducing stochasticity in the activity field enhances two- and multi-arm TDAP flexibility and leads to spontaneous formation of entangled clusters that are absent in deterministic systems.
(iii) Attractive cross-linkers attached to the tails of TDAP can stabilize dynamic aggregates of long TDAPs in high-activity regions, in particular when combined with temporally stochastic activity. The resulting structure, ranging from persistent asters to transient clusters or dispersed chains, depends on the interplay between chain length, propulsion strength, and linker affinity, which together determine whether polymers accumulate in high-, intermediate-, or low-activity regions.
}

Our findings could be experimentally tested using {microtubule gliding assays~\cite{eugene2025, catalano2025}}, synthetic swimmers~\cite{ji2019}.
Light patterns and light-controlled motor activation have also been used to generate and transport MT~\cite{ross2019, lemma2023} and actin~\cite{schuppler2016, zhang2021} assemblies such as asters, providing a potential method to explore the dynamics and stability of assemblies in inhomogeneous activity fields.




\textit{Acknowledgements.}---We thank Michael Lang for the discussions.
H.V. and A.S. acknowledge support from the Deutsche Forschungsgemeinschaft (DFG) under project numbers VA 2217/1-1 and SH 1275/5-1, respectively. J.U.S. acknowledges support from the Cluster of Excellence `Physics of Life' at TU Dresden.


\bibliography{references}

\foreach \x in {1,...,7}
{%
\clearpage
\includepdf[pages={\x}]{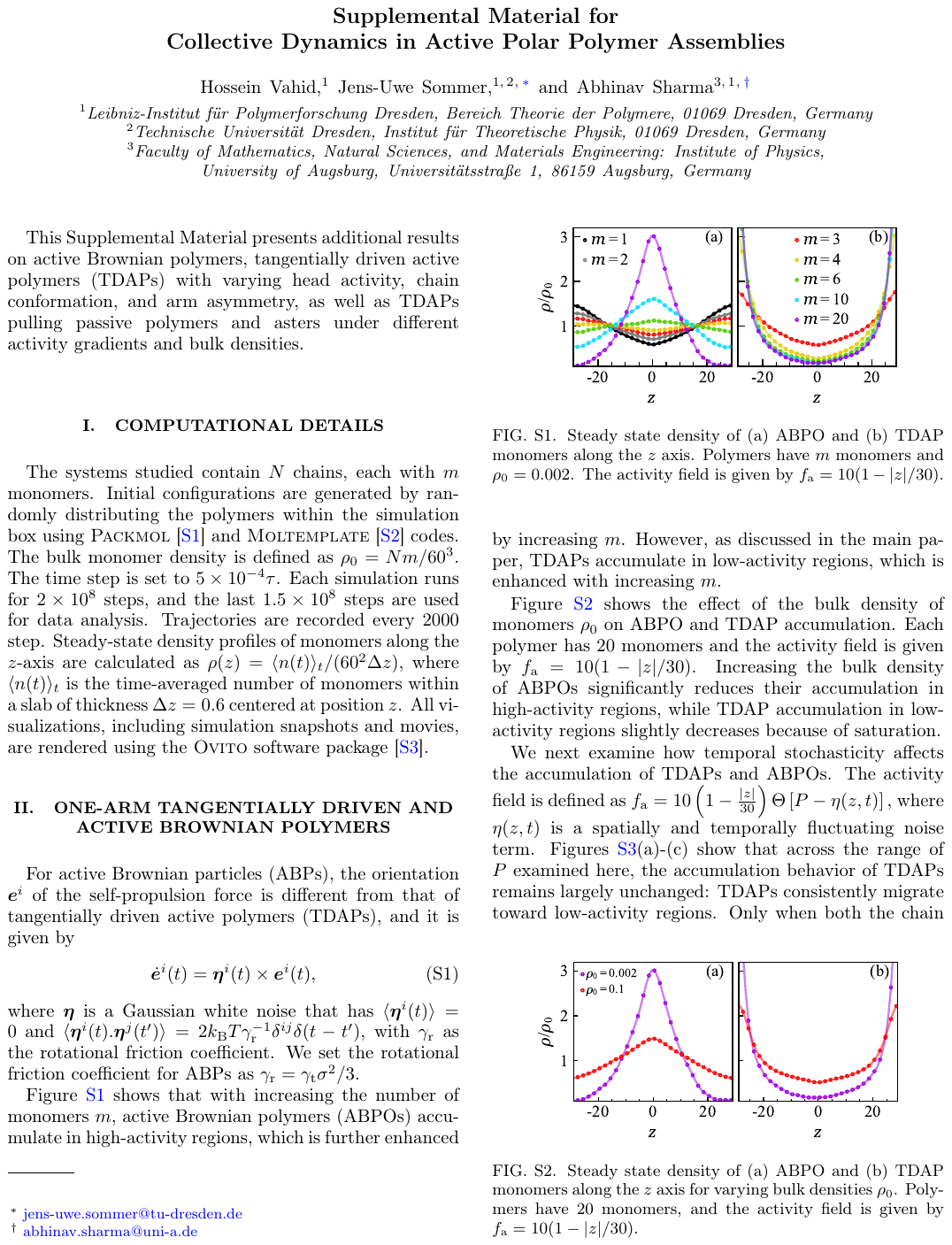} 
}

\end{document}